\documentclass[nofootinbib,showpacs,preprintnumbers,amsmath,amssymb,floatfix]
{revtex4-1}
\usepackage{graphicx}
\usepackage{wrapfig}
\usepackage{epstopdf}

\begin{document}


\title{Non-perturbative inputs for gluon distributions in the hadrons}

\vspace*{0.3 cm}

\author{B.I.~Ermolaev}
\affiliation{Ioffe Physico-Technical Institute, 194021}
 \author{S.I.~Troyan}
\affiliation{St.Petersburg Institute of Nuclear Physics, 188300
Gatchina, Russia}

\begin{abstract}
Description of hadronic reactions at high energies is conventionally done in the
framework of QCD factorization. All factorization convolutions comprise
non-perturbative inputs mimicking
non-perturbative contributions and perturbative evolution of those inputs. We construct
inputs for the gluon-hadron scattering amplitudes in the forward kinematics and, using
the Optical theorem, convert them into inputs for gluon distributions in the hadrons,
embracing the cases of polarized and unpolarized hadrons.
In the first place, we formulate mathematical criteria which any model for the inputs should obey and
then suggest a model satisfying those criteria.
This model is based on a simple reasoning: after emitting an active parton
off the hadron, the
remaining set of spectators becomes unstable and therefore it can be described through factors of the
resonance type, so we call it Resonance Model. We use it to obtain non-perturbative inputs for
gluon distributions in unpolarized and polarized hadrons for
all available types of QCD factorization: Basic, $K_T$- and Collinear Factorizations.
\end{abstract}


\maketitle

\section{Introduction}

\begin{figure}[h]
  \includegraphics[width=.15\textwidth]{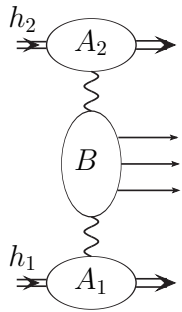}
  \caption{\label{gluonfig1} Amplitude for the Single-Parton Scattering of hadrons $h_1$ and $h_2$ , with active partons
being gluons. Blobs $A_{1,2}$ denote emission of the active gluons.
Interaction of those gluons is depicted by blob $B$, where the outgoing
arrows denote the produced partons. The outgoing double arrows on blobs $A_{1,2}$ stand for the final state
spectators.}
\end{figure}

\begin{figure}[h]
  \includegraphics[width=.35\textwidth]{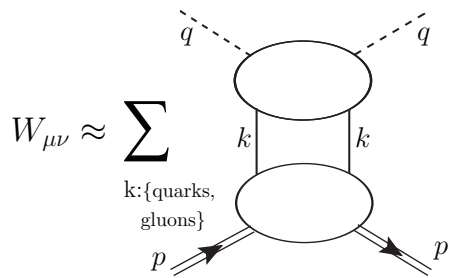}
  \caption{\label{gluonfig2} QCD factorization for
  the  DIS hadronic tensor. The $s$-cut
 of the graph is implied. The lowest blob includes the non-perturbative input while
 the upper blob corresponds to DIS off the active parton. }
\end{figure}

Description of hadronic reactions at high energies requires the use of QCD in the perturbative and
non-perturbative domains.
It is well-known that such calculations cannot be carried out in
the straightforward way because the non-perturbative QCD is poorly-known. The most efficient approximation approach
to theory of hadronic reactions is QCD factorization,
where perturbative QCD calculations are complemented by phenomenological fits or model expressions  which mimic
non-perturbative QCD contributions. Throughout the present paper we will address those expressions as
non-perturbative inputs. Technically speaking, such inputs act as initial conditions for
the evolution equations which account for perturbative contributions.
The inputs are supposed to approximately describe short-time dissociations
of each of the interacting hadrons into active partons and spectators.
Depending on the number of the active partons emitted by each hadron, there
can be Single-Parton Scattering (SPS) and Multi-Parton Scattering (MPS).
An example of the factorized amplitude for the
scattering of hadrons under the SPS approximation
is depicted in Fig.~\ref{gluonfig1}.
When such amplitudes are conjugated with the mirror amplitudes, the intermediate states
consist of two partons. For instance, QCD factorization of the
hadronic tensor $W_{\mu\nu}$ for DIS off a hadron with momentum
$p$ is represented in Fig.~\ref{gluonfig2}, where the $s$-cut is implied. The lowest blob
stands for a non-perturbative input describing emition/absorption of the active partons with momentum $k$ from the
initial hadron
 while the
upper blob corresponds to DIS off the active partons.

When the $s$-cut ($s = (p+q)^2$) in Fig.~\ref{gluonfig2} is not implied,
Fig.~\ref{gluonfig2} represents factorization of the
amplitude $A_{\mu\nu}$ of the elastic Compton scattering off the
hadron in the forward kinematics, with two-parton $t$-channel
intermediate states between the blobs.

The Optical theorem
relates the elastic scattering amplitudes in the forward kinematics to cross-sections.
 For example, it relates $A_{\mu\nu}$ to $W_{\mu\nu}$:

\begin{equation}\label{opt}
W_{\mu\nu} = \frac{1}{\pi} \Im_s A_{\mu\nu}.
\end{equation}

So, in order to calculate cross-sections or parton distributions under the SPS approximation,
it is sufficient to calculate the related elastic amplitudes with two-parton intermediate partons in the $t$-channel
only.
In order to avoid misunderstanding, we note that $t$-channel intermediate states inside the upper (perturbative) blob
can involve unlimited number of partons. Despite that there is a quite extensive literature on MPS scenario, the SPS scenario
is still the most popular. By this reason we focus on it in the present paper.

There are different kinds of QCD factorization in the
literature and each of them is
tailored to a specific perturbative
approach. For example, when DGLAP\cite{dglap} or its
generalization\cite{egtg1sum} to the small-$x$ region are used to
calculate $W_{\mu\nu}$,
the first step is to introduce an arbitrary mass scale $\mu$ acting as
a starting point of the perturbative $k^2_{\perp}$-evolution
from $\mu^2$ to $Q^2$, with $Q^2 \gg \mu^2$. This scale is called the factorization scale.
At the same moment, $\mu$
can act as a cut-off for infrared-divergent perturbative contributions.
In Collinear Factorization\cite{colfact}, transverse components of momentum $k$ in
Fig.~\ref{gluonfig2} are neglected, so these partons are regarded as collinear to the
incoming hadrons and therefore one-dimensional non-perturbative inputs
can be used.

 In contrast, BFKL\cite{bfkl} is free of
infrared divergences and because of that integrations over
transverse momenta in the BFKL ladders run down to zero.
This approach  operates
with essentially non-collinear initial partons, which excludes
neglecting their transverse components and as a result, it
excludes a simple
matching of BFKL with Collinear Factorization. By this reason,
$K_T$-Factorization\cite{ktfact} named also High-Energy
Factorization\cite{hefact} was suggested. These factorizations use
different ways of parametrization  of momenta $k$ of the connecting partons.
In Collinear
Factorization, the parametrization of $k$ is one-dimensional:
\begin{equation}\label{kcol}
k = \beta p,
\end{equation}
with $\beta$ being the longitudinal momentum fraction.
$K_T$-Factorization involves the same
longitudinal parameter $\beta$ and, in addition, accounts for the transverse momentum $k_{\perp}$:

\begin{equation}\label{kkt}
k = \beta p + k_{\perp}.
\end{equation}


There are different ways to construct the non-perturbative inputs in
Collinear and $K_T$- Factorizations.
Quite often, see e.g. Ref.~\cite{fits}, the inputs
applied in the context in Collinear Factorization
are introduced entirely on basis of practical reasons
without any theoretical grounds. In contrast, there also are the models with
solid theoretical background.
In particular, the models in Ref.~\cite{inputmodels} are actually based on
various theoretical approaches to approximate the problem of confinement:
the chiral quark-soliton model,
diquark model, etc. An overview of the most popular models of hadrons
can be found in Ref.~\cite{pasquini}.
The features of the saturation model\cite{golec}  are used in Refs.~\cite{jung,zotov}
for modeling the inputs in the context of $K_T$- Factorization while the model in Ref.~\cite{pumplin}
combines features of a variety of models for the Fock space wave function on the light cone.
Some features of this model are similar to the results obtained in Ref.~\cite{brod}.
Another interesting approach  is the lattice calculations.
They are a complementary method to study the
non-perturbative inputs for parton distributions.
Some of the recent results of the lattice calculations can be found is Ref.~\cite{lattice}.

Although the parametrization of momentum $k$  in $K_T$-
Factorization is more general than in Collinear Factorization,
it misses one more parameter for the longitudinal component of $k$,
 which can be seen from comparison of  Eq.~(\ref{kkt}) to
the standard Sudakov parametrization\cite{sud}:

\begin{equation}\label{sud}
k = - \alpha p' + \beta q' + k_{\perp},
\end{equation}
where the light-cone (i.e. $p'^2 = q'^2 = 0$) momenta $p',q'$ are made of the external momenta $p$ and $q$
satisfying the inequality $|pq| \gg |p^2|, |q^2|$
(for instance, they can be the momenta $p,q$ introduced in Fig.~2):
\begin{equation}\label{pq}
p' = p - x_p q,~q' = q - x_q p,~ x_p \approx p^2/w,~x_q \approx q^2/w,~w = 2p'q' \approx 2pq.
\end{equation}

In Ref.~\cite{egtfact} we presented a new kind of QCD factorization:  Basic Factorization.
It accounts for dependence of the factorized blobs on all components of momentum $k$ and because of that
it is the most general form of factorization. We proved that
Basic Factorization can step-by-step be reduced to $K_T$-Factorization and to
Collinear Factorization. Doing so, we
considered the non-perturbative inputs in the most general form, without
specifying them. In Ref.~\cite{egtfact} we inferred
general theoretical constraints on the non-perturbative inputs.
They follow from the
obvious requirement: despite the integrands in the factorization convolutions
for forward scattering amplitudes have  both infra-red (IR) and ultra-violet (UV) singularities,
integrations over $k$ in the factorization convolutions must yield finite results.
This is possible only if the non-perturbative inputs kill all divergences,
which leads to restrictions on the non-perturbative
inputs.
In Ref.~\cite{egtquark} we suggested a model for the non-perturbative inputs to
quark distributions in hadrons.
This model is based on a simple observation: after emitting an active parton
off the hadron, the set of remaining quarks and gluons (usually named spectators) becomes unstable, so it can be described by
expressions of the resonance type. Because of that we named this model in Ref,~\cite{egtquark} the Resonance Model.
We used the Resonance Model model so as to obtain the quark distributions in the hadrons whereas
the very important case of gluon distribution was left uninspected in Ref.~~\cite{egtquark}.

There is a certain similarity between handling quarks and gluons in Perturbative QCD. It allows us to believe that
main features of non-perturbative inputs for gluons and quarks can be alike. However,
the quark and gluon inputs cannot coincide. Firstly, they have different polarization structure:
the quark input should be a spinor whereas
the gluon one should be a tensor.
Secondly,
the well-known difference between the high-energy behavior of the gluon and quark
perturbative amplitudes can lead to an essential difference between the quark and gluon inputs.

  We think that this issue needs a thorough investigation and do it in the present paper.
In the first place we obtain restrictions on the gluon inputs which guarantee both IR and UV
stability of the factorization convolutions in Basic Factorization and then use those restrictions so as to
extend the Resonance Model to description of the non-perturbative inputs for
the gluon distributions in the hadrons. We consider here  both polarized and unpolarized hadrons.
Our strategy is to calculate the
gluon-hadron elastic scattering amplitudes in the forward kinematics and,
using the
Optical theorem, to
arrive at the gluon distributions in the hadrons in Basic Factorization. Then we reduce them
down to the expressions which can be used in $K_T$- and
Collinear Factorizations. As is well-known, $K_T$- Factorization (High-Energy Factorization)  by definition
can be used in the Regge (small-$x$) kinematics, so throughout the present paper we will consider the
parton distributions in this region only.

Our paper is organized as follows: In
Section II we study the elastic gluon-hadron amplitudes for the forward kinematic region in
the Born approximation, and then analyze the impact of radiative corrections. We investigate
the convergence of the factorization convolutions, using a general form for the non-perturbative
inputs, to determine constraints on
the non-perturbative inputs.
In Section III we formulate general criteria for the non-perturbative inputs in
Basic Factorization and use them to construct a Resonance Model.
Transition from the Basic Factorization to $K_T$- Factorization is considered
in Sect.~IV. In this Sect. we also compare our results with the ones available in the
literature.
In Sect.~V we reduce the inputs in $K_T$- Factorization down to the ones in Collinear Factorization.
We focus here on comparing our results with the standard DGLAP-fits.
At last, Sect.~VI is for our concluding remarks.

\section{Elastic gluon-hadron scattering amplitudes in the forward kinematics}

\begin{figure}[h]
  \includegraphics[width=.25\textwidth]{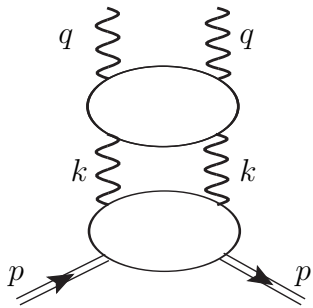}
  \caption{\label{gluonfig3} Factorization of amplitudes of hadron-gluon scattering in the
forward kinematics, with intermediate partons being gluons. The
upper blob corresponds to amplitudes of the elastic gluon scattering.}
\end{figure}

In this Sect. we consider
the elastic gluon-hadron amplitude $A$ in the forward kinematics and
inspect integrability of convolutions for
such amplitude in Basic Factorization. We study
the convolutions involving two-gluon intermediate $t$-channel states only because
we presume the approximation
of Single-Parton Scattering for the gluon distributions.
The convolutions describing the gluon-hadron amplitude are depicted by the sum of two
graphs\footnote{Throughout the paper we consider the $t$-channel  color singlets only.}:
the graph
 in Fig.~\ref{gluonfig3} and
a similar graph, where $q$ is replaced by $-q$.

Our aim here is to obtain mathematical restrictions on the non-perturbative inputs. In order to do it,
we use the obvious reasoning: Integration over momentum $k$ in the factorization convolutions
(see Fig.~\ref{gluonfig3}) runs over the whole phase space and it should yield a finite result.
Features of perturbative components of the convolutions are known, at least qualitatively,
so integrability requirement can be used to establish necessary restrictions on the non-perturbative
inputs in a general form. Then any model for the inputs should  meet these restrictions. Therefore, the
restrictions can
be used as criteria for accepting or rejecting the models.
To begin with, we consider amplitude $A$ in the Born approximation and then examine an impact of the radiative
corrections. We show that,
in contrast to the concepts of $K_T$- and Collinear Factorizations,
Basic Factorization allows the simple scenario, where the upper blob in
 in Fig.~\ref{gluonfig3} can be regarded as altogether perturbative one
 while the lowest blob includes non-perturbative
contributions only.

\subsection{Gluon-hadron scattering amplitudes in the Born approximation}

\begin{figure}[h]
  \includegraphics[width=.25\textwidth]{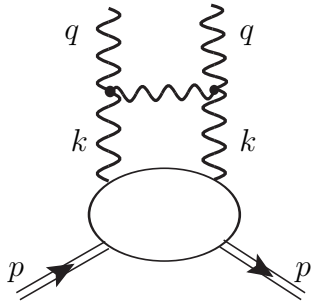}
  \caption{\label{gluonfig4} Factorization  for amplitude of hadron-gluon scattering
  in the Born approximation.}
\end{figure}

Let us consider the simplest case of the factorization convolutions for the gluon-hadron amplitude, where the non-perturbative
 input is convoluted with the gluon-gluon perturbative amplitude
in the Born approximation.
In this case the upper blob in Fig.~3 is represented by two simple graphs, each with the single-gluon exchange.
The first graph, with a non-zero imaginary part in $s$, is depicted
in Fig.~\ref{gluonfig4} and the second graph, with a non-zero imaginary part in $u$, can be obtained from the first one by
replacement $q \to - q$.

A remarkable feature of Basic Factorization is that the analytic expression
corresponding to the convolution graphs can be obtained with the use of the
standard Fetnman rules.
Applying these rules to the graph in Fig.~\ref{gluonfig4} and complementing the
result by the exchange $q \to -q$,  we arrive at the following
expression for the Born gluon-hadron amplitude $A^B$
in Basic Factorization:

\begin{equation}\label{aborn}
A^B= \frac{4 \pi \alpha_s N}{(2 \pi)^4}
\int d^4 k  ~l_{\mu} l'_{\nu} \left[\frac{H_{\mu\nu\lambda \rho}(q,k)}{ s_2 + \imath \epsilon}
+ \frac{H_{\nu\mu\lambda \rho}(-q,k)}{u_2 + \imath \epsilon} \right] \frac{1}{k^2 k^2}
T_{\lambda \rho} (k,p, S),
\end{equation}
where

\begin{equation}\label{s2}
s_2 = (q+k)^2,~ u_2 = (q-k)^2,
\end{equation}

$N=3$ is the color factor and $S$ denotes the hadron spin.
The term in squared brackets corresponds to the Born amplitude for
gluon-gluon scattering, $l_{\mu}(q)$ and $l'_{\nu}(q)$ denote the polarization vectors of the
external gluons,
the terms $1/k^2$ correspond to propagators of the connecting gluons
and the input $T_{\lambda \rho}$ contains non-pertrurbative contributions
only. Throughout the paper we use the Feynman gauge. Let us notice that both $T_{\lambda\rho}$ and the term in the squared
brackets are dimensionless.
The perturbative term $H_{\mu\nu\lambda \rho}$ in Eq.~(\ref{aborn}) is

\begin{eqnarray}\label{h}
H_{\mu\nu\lambda \rho} = \left[(-q -2k)_{\mu} g_{\lambda\sigma} +(2q + k)_{\lambda} g_{\mu\sigma}
+ (k-q)_{\sigma} g_{\mu\lambda}\right]
\left[(q+2k)_{\nu} g_{\rho\sigma} + (-2q - k)_{\rho} g_{\nu\sigma} +(q-k)_{\sigma}g_{\nu\rho}\right].
\end{eqnarray}

We remind that in the present paper we consider the gluon-hadron scattering amplitudes and
parton distributions in the small-$x$ region. The point is that we are going to
reduce Basic Factorization to $K_T$- Factorization which is defined in the small-$x$ region only.
It is easy to check that $A^B$ with $H_{\mu\nu\lambda \rho}$ given by Eq.~(\ref{h}) is gauge-invariant in the small-$x$,
though approximately\footnote{The gauge invariance for the quark-hadron scattering amplitudes
was considered in Ref.~\cite{egtquark}.}

In order to make further progress we should specify $T_{\lambda\rho}$.
First of all, we should fix its polarization (tensor) structure. For instance,
when the hadron and the gluons are unpolarized, $T_{\lambda\rho}$ should be symmetric in $\lambda, \rho$.
The general form of such a tensor, $T^{(gen)}_{\lambda\rho}$ is
\begin{equation}\label{tgen}
T^{(gen)}_{\lambda\rho} = g_{\lambda\rho} A + p_{\lambda}p_{\rho}B + (p_{\lambda}k_{\rho} +
k_{\lambda}p_{\rho})C + k_{\lambda}k_{\rho}D,
\end{equation}
but it involves four arbitrary invariant amplitudes $A,..,D$. Trying to diminish their number,
we recall an instance from perturbative QCD, namely that,
when the hadron is replaced by a bare quark, $T_{\lambda\rho}$ is replaced by the quark-gluon
Born amplitude $Q_{\lambda\rho}$. It consists of the unpolarized part $Q^U_{\lambda\rho}$
and the spin-dependent part $Q^S_{\lambda\rho}$:

\begin{equation}\label{qussum}
Q_{\lambda\rho} = Q^U_{\lambda\rho} + Q^S_{\lambda\rho},
\end{equation}
with
\begin{eqnarray}\label{qus}
Q^U_{\lambda\rho} &=& \left(2 p_{\lambda}p_{\rho} - k_{\lambda}p_{\rho}+ p_{\lambda}k_{\rho} - pk~g_{\lambda\rho}\right)
\left(\frac{- 8 \pi\alpha_s C_F}{(p -k)^2 - m^2_q + \imath \epsilon}\right),
\\ \nonumber
Q^S_{\lambda\rho} &=& \imath m_q\epsilon_{\lambda\rho\sigma\tau}k_{\sigma}(S_q)_{\tau}
 \left(\frac{- 8 \pi\alpha_s C_F}{(p -k)^2 - m^2_q + \imath \epsilon}\right),
\end{eqnarray}
where $m_q$ is the quark mass, $S_q$ is the quark spin and $C_F = 4/3$.
We suggest the simplest generalization of Eqs.~(\ref{qussum},\ref{qus}), for the hadrons:
we presume that $T_{\lambda\rho}$  keeps the polarization structure of $Q_{\lambda\rho}$. It means that

\begin{equation}\label{tussum}
T_{\lambda\rho} = T^U_{\lambda\rho} + T^S_{\lambda\rho},
\end{equation}
with

\begin{eqnarray}\label{tus}
T^U_{\lambda\rho} &=& \left(2 p_{\lambda}p_{\rho} - k_{\lambda}p_{\rho} -
p_{\lambda}k_{\rho} + pk~g_{\lambda\rho}\right)
M_U (s_1, k^2),
\\ \nonumber
T^S_{\lambda\rho} &=& \imath M_h\epsilon_{\lambda\rho\sigma\tau}k_{\sigma}S_{\tau}
M_S (s_1, k^2),
\end{eqnarray}
where $M_h, S$ are the hadron mass and spin respectively, and

\begin{equation}\label{s1}
s_1 = (p-k)^2.
\end{equation}
 The invariant amplitudes
$M_U$ and $M_S$  may either coincide or differ from each other. Substituting $T^{U,S}_{\lambda\rho}$ in
Eq.~(\ref{aborn}), we obtain:

\begin{equation}\label{abu}
A^B_U = \frac{4 \pi \alpha_s N}{(2 \pi)^4}
\int \frac{d^4 k}{k^2 k^2}  \left[
\frac{N^{(1)}_s}{s_2 + \imath \epsilon}  +
\frac{N^{(1)}_u}{u_2 + \imath \epsilon} \right] M_U (s_1,k^2),
\end{equation}

\begin{equation}\label{abs}
A^B_S = \frac{4 \pi \alpha_s N}{(2 \pi)^4}
\int \frac{d^4 k}{k^2 k^2}  \left[
\frac{N^{(2)}_s}{s_2 + \imath \epsilon}  +
\frac{N^{(2)}_u}{u_2 + \imath \epsilon} \right]M_S (s_1,k^2),
\end{equation}
with

\begin{eqnarray}\label{n12}
N^{(1)}_s &=& 4 k^2(2m^2_h + 2pk) -16(2 (pq)^2 +2 (pq)(pk) - q^2 (pk))
+u_2 (2 M^2_h - 2 pk),
\\ \nonumber
N^{(2)}_s &=&
 \imath M_h \epsilon_{\lambda\rho\sigma\tau} S_{\tau}
\left[8q_{\rho} \left(l'_{\lambda}k_{\sigma}(kl) -  l_{\lambda}k_{\sigma}(kl')\right)
- u_2 k_{\sigma}l_{\lambda}l'_{\rho}\right].
\end{eqnarray}

We have done summation over the gluon polarizations in the expression for $N^{(1)}_s$ in Eq.~(\ref{n12}).
In what follows we will use the Sudakov variables defined in Eq.~(\ref{sud}). In their terms

\begin{equation}\label{sudinv}
2pk = - \alpha w + \beta x_p w,~~2qk = \beta w - \alpha x w,~
k^2 = - \alpha\beta w + k^2_{\perp}.
\end{equation}

\subsection{Analysis of IR and UV singularities of the Born amplitudes}

Let us consider integrability of the Born amplitudes in Eqs.~(\ref{abu},\ref{abs}).
Integration over momentum $k$ in both convolutions runs over the whole phase space and
result of the integration must be finite despite the integrands may have singularities
which must be regulated.
First of all, let us note that the expressions in parentheses  in Eqs.~(\ref{abu},\ref{abs})
stand for the gluon perturbative amplitudes in the Born approximation, which
are obviously free of divergences. In contrast, the factors $1/k^2$ in Eqs.~(\ref{abu},\ref{abs})
become divergent at $k^2 = 0$.
We cannot implement
any IR cut-off to regulate those singularities because
there is no physical reason to restrict the integrations regions.
It leaves us with the only option:
amplitudes $M_{U,S}$ in Eqs.~(\ref{abu},\ref{abs}) at small $k^2$ must decrease
rapidly enough in order to kill these IR singularities.
Therefore we infer that $M_{U,S}$  should satisfy the following restrictions
at $k^2 \to 0$:

\begin{equation}\label{mir}
M_{U,S}(s_1,k^2)  \sim \left(k^2\right)^{1 + \eta},
\end{equation}
with $\eta > 0$. Eq.~(\ref{mir}) sets  IR stability for the integrals in Eqs.~(\ref{abu},\ref{abs}).
Then, there may be UV singularities in Eqs.~(\ref{abu},\ref{abs}) at large $|k|$,
i.e. at large $|\alpha|$  in terms of the Sudakov variables, when the integration over $\alpha$ goes first.
It is easy to see that the integrands in Eqs.~(\ref{abu},\ref{abs}) at large $|\alpha|$
behave as

\begin{equation}\label{uv}
\sim \frac{\alpha^2}{\alpha^3} M_{U,S}
\end{equation}
which can be UV-divergent. In order to kill this divergence, $M_{U,S}$ should decrease at large $\alpha$:

\begin{equation}\label{muv}
M_{U,S} \sim \alpha^{- \chi},
\end{equation}
with $\chi > 0$. In other words, the non-perturbative inputs should decrease, when their
invariant energy $s_1$ grows. Also, we note that the restrictions in Eqs.~(\ref{mir},\ref{muv})
are valid at any choice of
the polarization structure of $T_{\lambda\rho}$, including the most general parametrization
in Eq.~(\ref{tgen}).

\subsection{Gluon-hadron scattering amplitudes beyond the Born approximation}

Diagrammatically, accounting for the radiative corrections to the Born amplitudes $A^B_{U,S}$ of Eqs.~(\ref{abu},\ref{abs})
means adding more quark and gluon propagators
to the single-gluon exchange in Fig.~4 and summing up
contributions of such graphs.
This procedure converts the Born perturbative amplitudes of Eqs.~(\ref{abu},\ref{abs})
into amplitudes $A^{(pert)}_{U,S}$ (the upper blob in Fig.~3) but keeps unchanged the non-perturbative inputs $M_{U,S}$
(the lowest blob in Fig.~3).
Such natural separation of the perturbative and non-perturbative contributions is a remarkable
feature of our approach and perfectly agrees with the factorization concept.

Replacing the Born amplitudes in Eqs.~(\ref{abu},\ref{abs}) by the amplitudes $A^{(pert)}_{U,S}$, we arrive at
the gluon-hadron amplitudes $A_U, A_S$ beyond the Born approximation:

\begin{equation}\label{aus}
A_{U,S} =
\int \frac{d^4 k}{k^2 k^2}  A^{(pert)}_{U,S} \left(s_2/q^2, s_2/k^2\right)  M_{U,S} (s_1,k^2).
\end{equation}
Applying the Optical theorem to Eq.~(\ref{aus}), we obtain the spin-independent, $D_U$, and polarized, $D_S$, distributions
of gluons in the hadrons:

\begin{equation}\label{dus}
D_{U,S} =
\int \frac{d^4 k}{k^2 k^2}  D^{(pert)}_{U,S} \left(s_2/q^2, s_2/k^2\right)  G_{U,S} (s_1,k^2),
\end{equation}
where
\begin{equation}\label{dg}
D^{(pert)}_{U,S} (s_2,q^2,k^2) = \Im_{s_2} A^{(pert)}_{U,S},  ~~G_{U,S} (s_1,k^2) = \Im_{s_1} M_{U,S} (s_1,k^2).
\end{equation}
We remind that the invariant sub-energies $s_{1,2}$ are defined as follows: $s_1 = (q + p)^2, s_2 = (k - p)^2$.

Now we prove that the difference between perturbative amplitudes $A^{(pert)}_{U,S}$
and their
Born values does not changes the conditions (\ref{mir},\ref{muv}) of IR and UV stability.
Before doing it, let us notice that $A^{(pert)}_{U,S}$ depends on $s_2, k^2, q^2$, so, being dimensionless, it can be parameterized,
for instance, as follows:

\begin{equation}\label{apertus}
A^{(pert)}_{U,S} (s_2, q^2, k^2) =
A^{(pert)}_{U,S} \left(s_2/q^2, s_2/k^2\right).
\end{equation}

Perturbative higher-loop contributions can induce both IR and UV divergences in $A^{(pert)}_{U,S}$.
All those divergences should be regulated before using $A^{(pert)}_{U,S}$ in the factorization convolutions (\ref{aus}).
We start with
considering regularization of  IR divergences in $A^{(pert)}_{U,S}$. They appear from integrations
of gluon loops over
the regions,
where some of the virtual gluons are soft. In addition, the problem of IR divergences is essential
for soft quarks when their masses
are neglected. The ways of regulating IR divergences are different for the amplitudes with
on-shell and off-shell external partons. We will address such amplitudes as the on-shell and off-shell amplitudes respectively.
For the on-shell amplitudes, the IR divergences
are regulated by suitable IR cut-offs, while virtualities of the external partons play the role of
natural IR cut-offs for the off-shell amplitudes.
 The latter was shown first in Ref.~\cite{sud} and
then was used in numerous calculations. Therefore, $q^2$ and
$k^2$ can be used as IR cut-offs for $A^{(pert)}_{U,S}$.

Now let us consider regulation of the UV divergences in $A^{(pert)}_{U,S}$. This problem is solved easily.
The point is that QCD is a
renormalizable theory, so all UV divergences are automatically absorbed by appropriate
redefinitions of $\alpha_s$ and masses of involved quarks and gluons.

Unfortunately, regulating IR singularities of $A^{(pert)}_{U,S}$ does not eliminate the IR divergences of the factorization
convolutions (\ref{aus}).
Indeed,
the  IR-sensitive contributions in $A^{(pert)}_{U,S}$ are $\sim \ln^n (w \beta/k^2)$ (or
$\sim \ln^n (q^2/k^2)$), so these contributions are IR stable at $k^2 \neq 0$ but become divergent at the point
 $k^2 = 0$. Integration in Eqs.~(\ref{aus})
covers the whole phase space, so the integration region includes the point $k^2 = 0$.
Such logarithmic divergences were addressed as mass singularities in
the pioneer studies\cite{colfact}  of QCD hard processes in the framework of Collinear Factorization, where the mass singularities were
moved from the hard part to universal structure functions.
However, such a treatment of the mass singularities becomes
unproductive in the Regge (small-$x$) kinematics which we pursuit throughout this paper.
Furthermore, the logarithmic singularities are the only kind of IR divergences
in Collinear Factorization but in Basic Factorization we should handle both
them and  the power singularity $\sim 1/(k^2 k^2)$
(see Eqs.~(\ref{aus})). In Sect.~IIB we demonstrated that the power singularity
$1/(k^2 k^2)$ can be regulated by the condition (\ref{mir}).
Obviously, this condition persists when the singularity $1/(k^2 k^2)$ is accompanied by the
logarithmically divergent  terms.  It is worth noticing here that the well-known power factor $s/k^2$
in the spin-independent amplitude
appears in the lowest order in $\alpha_s$, borrowing $k^2$ from the denominator in Eqs.~(\ref{aus}),
so its impact on IR stability was taken into account in Eq.~(\ref{mir}).

Finally, it is easy to prove that accounting for the perturbative radiative corrections does not change
the condition (\ref{muv}) of UV stability of the factorization convolutions
at large $|\alpha|$.  Indeed, the fact that the perturbative QCD is renormalizable means  that
the fastest-growing with energy terms in amplitudes $A^{(pert)}_{U,S}$  are of logarithmic
kind, i.e.
 $\sim \ln^n \left(w \alpha/q^2\right), \ln^n \left(w \alpha/k^2\right)$,
so such logarithmic divergences can be regulated by
the power restriction in Eq.~(\ref{muv}).

\section{Modeling the non-perturbative invariant amplitudes $M_{U,S}$}

In this Sect. we consider first the Resonance Model for non-perturbative amplitudes $M_{U,S}$ and then proceed to
the non-perturbative inputs for the
gluon distributions. Our models is based on simple and clear-cut physical arguments
which we list below and call them criteria which any model for $M_{U,S}$ should satisfy.
Although constructing such a model cannot be done unambiguously, any proposed model should
satisfy the following criteria:

\subsection{General criteria for modeling non-perturbative inputs}

\textbf{Criterion} \textbf{(i)} Expressions for $M_{U,S}$ should satisfy the requirements of
IR and UV stability of the factorization convolutions given in Eqs.~(\ref{mir},\ref{muv}). \\

\textbf{Criterion} \textbf{(ii)} Expressions for $M_{U,S}$ should have non-zero imaginary parts in $s_1$ to make
possible the use of the Optical theorem. It is necessary in order to proceed from elastic gluon-hadron amplitudes to
gluon distributions in the hadrons.  \\

\textbf{Criterion} \textbf{(iii)} Expressions for $M_{U,S}$ should enable the step-by-step reduction of Basic Factorization
down to $K_T$- and Collinear Factorizations. Such a reduction was suggested in
general in Ref.~\cite{egtfact}.

\subsection{Specifying non-perturbative gluon inputs in Basic Factorization}

We suggest the following model expressions for the amplitudes $M_{U,S}(s_1, k^2)$ :

\begin{equation}\label{musgen}
M_U(s_1, k^2) = R_U\left(k^2\right) Z_U \left(s_1\right),~~M_S(s_1, k^2) = R_S\left(k^2\right) Z_S \left(s_1\right).
\end{equation}
Eq.~(\ref{musgen}) reads that dependences on $s_1$ and $k^2$ are separated. This is done for the sake
of simplicity. We could assume that $R = R (k^2, s_1)$ but we find it unnecessary.
As handling $M_U$ and $M_S$ is much the same, we skip the subscripts $U,S$ and will
work with the
generic notations

\begin{equation}\label{mrz}
M(s_1, k^2) = R(k^2) Z(s_1)
\end{equation}
instead of Eq.~(\ref{musgen}). In order to satisfy
Eq.~(\ref{mir}), $R$ in Eq.~(\ref{mrz}) should exhibit such a behavior:

\begin{equation}\label{rsmallk}
R \sim \left(k^2\right)^{1 + \eta}
\end{equation}
at small $k^2$. According to Eq.~(\ref{muv}), $RZ \sim \alpha^{- \chi}$ at large $|\alpha|$.
A behavior of $R$ at large $|\alpha|$ is unknown. However, it can be that the small-$k^2$ behavior
at large $|\alpha|$ is again $R \sim (k^2)^{ 1 + \eta}$, i.e. according to Eq.~(\ref{sudinv})

\begin{equation}\label{rbiga}
R \sim \alpha^{ 1 + \eta}.
\end{equation}

This is the most UV divergent case which should not be ignored,
so in order to satisfy Eqs.~(\ref{muv},\ref{rbiga}),
$Z$ should behave as follows: at large $|\alpha|$

\begin{equation}\label{zuv}
Z \sim \alpha^{-1 -\eta - \chi }.
\end{equation}
If $R$ does not grow at large $|\alpha|$, we can all the same use $Z$ obeying Eq.~(\ref{zuv}).
In order to specify $Z$, we suggest the guiding idea:
After emitting an active parton by the initial hadron, the remaining set of quark and gluons pick up a color and
therefore it cannot be stable. Unstable states, in general, are known to be often modeled by expressions of the resonance type. In
particular, we can approximate $Z$ in Eq.~(\ref{mrz}) by the following expression:

\begin{equation}\label{zgen}
Z (s_1) \approx Z_n (s_1) = \prod_{r = 1}^{n} \frac{1}{\left(s_1 - M^2_r + \imath \Gamma_r \right)},
\end{equation}
with arbitrary positive integer $n$.
In order to get $Z_n(s_1)$ satisfying Eq.~(\ref{zuv}), we should choose $n > 1$.  For the sake of simplicity,
we consider the minimal value $n = 2$.
Obviously, $Z_2$ can also be written as an interference of two resonances:

\begin{equation}\label{zres}
Z (s_1) = \frac{1}{\left(\Delta M^2_{12} + \imath \Delta \Gamma_{12}\right)}
\left[\frac{1}{\left(s_1 - M^2_1 + \imath \Gamma_1\right)} -
\frac{1}{\left(s_1 - M^2_2 + \imath \Gamma_2\right)}
\right],
\end{equation}
with $\Delta M^2_{12} = M^2_1 - M^2_2,~  \Delta \Gamma_{12} = \Gamma_1 - \Gamma_2$.
In terms of the Sudakov variables,

\begin{equation}\label{msud}
M (w\alpha, k^2) = \frac{R(k^2)}{C_Z}
\left[\frac{1}{\left(w \alpha - \mu^2_1 + k^2 + \imath \Gamma_1\right)} -
\frac{1}{\left(w \alpha - \mu^2_2 + k^2 + \imath \Gamma_2\right)}
\right],
\end{equation}
where $C_Z = \left(\Delta M^2_{12} + \imath \Delta \Gamma_{12}\right)$ and
$\mu^2_{1,2} = M^2_{1,2} - p^2$.
Applying the Optical theorem to Eq.~(\ref{msud})
allows us to obtain the non-perturbative contribution $G$ to the gluon distributions in the hadrons:

\begin{equation}\label{d}
G = -\Im_{s_1} T =
 \frac{R(k^2)}{C_Z}
\left[\frac{\Gamma_1}{\left(w \alpha + k^2 - \mu^2_1\right)^2 +\Gamma_1^2} -
\frac{\Gamma_2}{\left(w \alpha + k^2 - \mu^2_2 \right)^2 +\Gamma_2^2}\right].
\end{equation}

Obviously, this expression
is of the Breit-Wigner type.

\section{Non-perturbative gluon inputs in $K_T$- Factorization}

The expression for the gluon-hadron scattering amplitude $M(\alpha,\beta,k^2)$  in Eq.~(\ref{msud})
corresponds to Basic Factorization.
In order to reduce $M(\alpha,\beta,k^2)$  to the amplitude $M_{KT}(\beta,k^2_{\perp})$ of
the same process in
$K_T$-Factorization,
we should integrate out the $\alpha$-dependence of $M(\alpha, \beta, k^2)$. In principle, as soon as
However, in factorization convolutions (see e.g. Figs.~2,3,4 ) both the upper and the lowest blobs depend on
$\alpha$, so  one cannot integrate $M(\alpha, \beta, k^2)$ only, without integration of the perturbative blob.
Therefore,
$M(\beta, k^2_{\perp})$ cannot be derived from $M(\alpha, \beta, k^2)$
in the straightforward way, which makes impossible a straightforward reduction of
Basic Factorization to $K_T$-Factorization. Nevertheless, it can be done approximately.
The perturbative blobs (the upper blob in Figs.~2,3,4) approximately do not depend on $\alpha$
in the region

 \begin{equation}\label{abk}
 w |\alpha\beta| \ll k^2_{\perp}
 \end{equation}

and the non-perturbative blobs in this region are independent of $\beta$. Indeed, the upper
blob depends on $\alpha$ through $k^2$ only and in
the region (\ref{abk}) $k^2 \approx - k^2_{\perp}$. At the same time, the region
(\ref{abk}) is known to be the region for gaining  the perturbative contributions most essential
at small $x$.
For instance, all BFKL logarithms come from  this region.
So, Eq.~(\ref{abk}) makes possible
to integrate  $M \left(\alpha,\beta, k^2\right)$ independently of perturbative contributions.
It is convenient to define the non-perturbative input $M_{KT}$ in $K_T$ Factorization as follows:

\begin{equation}\label{tktgen}
k^2_{\perp} M_{KT} (\beta,k^2_{\perp} ) = \int^{\alpha_0}_{-\alpha_0} d \alpha M(\alpha,\beta, k^2),
\end{equation}
with

\begin{equation}\label{azero1}
\alpha_0 \ll k^2_{\perp}/(w \beta).
\end{equation}

%
%
So, Eq.~(\ref{azero1})
is satisfied when

\begin{equation}\label{azero}
\alpha_0 = \xi k^2_{\perp}/\beta,
\end{equation}
with a positive $\xi$ obeying the inequality $\xi \ll 1$.
Combining Eqs.~(\ref{msud},\ref{tktgen},\ref{azero})
 we arrive at the following expression for the non-perturbative input
$M_{KT}$ in $K_T$ -Factorization:

\begin{eqnarray}\label{mkt}
M_{KT} &\approx& \frac{R\left(k^2_{\perp}\right)}{k^2_{\perp}}
\left[
\frac{1}{\xi k^2_{\perp}/\beta - \mu^2_1 +\imath \Gamma_1}
+ \frac{1}{\xi k^2_{\perp}/\beta - \mu^2_2 +\imath \Gamma_2}
+ \frac{1}{\xi k^2_{\perp}/\beta + \mu^2_1 -\imath \Gamma_1}
+ \frac{1}{\xi k^2_{\perp}/\beta + \mu^2_2 -\imath \Gamma_2}\right]
\\ \nonumber
&=& R_{KT}\left(k^2_{\perp}\right)
\left[
\frac{1}{k^2_{\perp}/\beta - {\mu'}^2_1 +\imath \Gamma'_1}
+ \frac{1}{k^2_{\perp}/\beta - {\mu'}^2_2 +\imath \Gamma'_2}
+ \frac{1}{k^2_{\perp}/\beta + {\mu'}^2_1 -\imath \Gamma'_1}
+ \frac{1}{k^2_{\perp}/\beta + {\mu'}^2_2 -\imath \Gamma'_2}\right],
\end{eqnarray}
where $R_{KT} = R/(\xi k^2_{\perp} C_Z)$
and
\begin{equation}\label{muprime}
{\mu'}^2_{1,2} = \mu^2_{1,2}/\xi,~\Gamma'_{1,2} = \Gamma_{1,2}/\xi .
\end{equation}

We also used the well-known observation that the most essential
region in the factorization convolutions is the small-$\beta$ region,
where $\xi/\beta$ is not far from $x$.
Eq.~(\ref{mkt}) is valid when

\begin{equation}\label{azeromg}
w \alpha_0 \gg \mu^2_{1,2}, \qquad \Gamma_{1,2} \gg \Delta \mu^2, \Delta \Gamma,
\end{equation}
i.e. when $k^2_{\perp}/\beta$ is pretty far from the resonance region
$k^2_{\perp}/\beta \sim \mu'^2_{1,2}$. The closer
$k^2_{\perp}/\beta$ is to $\mu'^2_{1,2}$ the greater are corrections to Eq.~(\ref{mkt}).
It means that transition from Basic Factorization to $K_T$-Factorization considered
in Ref.~\cite{egtfact} is straightforwardly done
outside the resonance region. After that, Eq.~(\ref{mkt}) can be  analytically continued
in the resonance region.
As a result,
we arrive at the gluon-hadron amplitude $A_{KT}$  and the gluon distribution $D_{KT}$ in $K_T$-Factorization:

\begin{equation}\label{akt}
A_{KT} = \int \frac{d \beta}{\beta} \frac{d k^2_{\perp}}{k^2_{\perp}} A^{(pert)}_{KT} (w \beta,q^2,k^2_{\perp}) M_{KT} (\beta, k^2_{\perp})
\end{equation}

\begin{equation}\label{dkt}
D_{KT} = \int \frac{d \beta}{\beta} \frac{d k^2_{\perp}}{k^2_{\perp}} D^{(pert)}_{KT} (w \beta,q^2,k^2_{\perp}) G_{KT} (\beta, k^2_{\perp}),
\end{equation}
where $A^{(pert)}_{KT}$  and $D^{(pert)}_{KT} = \Im A^{(pert)}_{KT}$ are the perturbative contributions whereas $M_{KT}$ and
$G_{KT} = \Im M_{KT}$ (cf. Eq.~(\ref{dg})) are the non-perturbative inputs. It is convenient to represent
$G_{KT}$ in the following form:

\begin{equation}\label{grb}
G_{KT} = G_R + G_B,
\end{equation}
with
\begin{equation}\label{gr}
G_R  = R_{KT}\left(k^2_{\perp}\right)
\left(
\frac{\Gamma'_1}{(k^2_{\perp}/\beta - {\mu'}^2_1)^2 +{\Gamma'}_1^2}
+ \frac{\Gamma'_2}{(k^2_{\perp}/\beta - {\mu'}^2_2)^2 + {\Gamma'}_2^2} \right)
\end{equation}
and

\begin{equation}\label{gb}
G_B = R_{KT}\left(k^2_{\perp}\right) \left(\frac{\Gamma'_1}{(k^2_{\perp}/\beta + {\mu'}^2_1)^2 + {\Gamma'}_1^2}
+ \frac{\Gamma'_2}{(k^2_{\perp}/\beta + {\mu'}^2_2)^2 + {\Gamma'}_2^2}\right).
\end{equation}
$G_R$ and $G_B$ obviously consist of expressions of the Breit-Wigner type.
 Signs of ${\mu'}^2_1$ and ${\mu'}^2_2$ cannot be fixed a priory, but
$G_R  \leftrightarrows G_B$ when ${\mu'}^2_{1,2} \to - {\mu'}^2_{1,2}$,
so we can consider the case of positive ${\mu'}^2_{1,2}$ without loss of generality.
We remind that $k^2_{\perp}/\beta > 0$, so $G_R$ is within the
resonant region $k^2_{\perp}/\beta \sim \mu'^2_{1,2}$  while $G_B$ is out of that region. Eqs.~(\ref{gr},\ref{gb})
allows us to represent the gluon distribution
$D_{KT}$ in $K_T$ -Factorization  as the sum of  its
resonance part $D_{KT}^R$ and the background contribution $D_{KT}^B$:

\begin{equation}\label{dktrb}
D_{KT} = D_{KT}^R + D_{KT}^B,
\end{equation}
with

\begin{equation}\label{dktr}
D_{KT}^R = \int \frac{d \beta}{\beta} \frac{d k^2_{\perp}}{k^2_{\perp}} D^{(pert)}_{KT} (w \beta,q^2,k^2_{\perp}) G_R (\beta, k^2_{\perp})
\end{equation}
and
\begin{equation}\label{dktb}
D_{KT}^B = \int \frac{d \beta}{\beta} \frac{d k^2_{\perp}}{k^2_{\perp}} D^{(pert)}_{KT} (w \beta,q^2,k^2_{\perp}) G_B (\beta, k^2_{\perp}).
\end{equation}

Representation of $D_{KT}$ through the resonance $D_{KT}^R$ and background $D_{KT}^B$ contributions is similar to the structure of expressions
in the Duality concept.

\subsection{Discussion of the distribution $\textbf{G}_{\textbf{KT}}$}

First of all, let us compare the mass parameters $\mu_{1,2}$
of the gluon distribution $D$ of Eq.~(\ref{d})
in Basic Factorization and the parameters $\mu'_{1,2}$
of the gluon distribution $D_{KT}$ of Eq.~(\ref{dkt})
in $K_T$- Factorization. They are related by Eq.~(\ref{muprime}).
As $\xi \ll 1$, Eq.~(\ref{muprime}) means that $\mu'_1 \gg \mu_1,~\mu'_2 \gg \mu_2$. It means that
despite the non-perturbative input $G$ is defined in Basic Factorization within the non-perturbative domain,  the
resonance maximums of $G_{KT}$ are located at perturbative values of
$k^2_{\perp}/\beta > 0$.
Then, we would like to comment on the factor $R_{KT}$ in Eqs.~(\ref{mkt},\ref{gr},\ref{gb}). This factor
describes a dependence of $G_{KT}$ on $k^2_{\perp}$.
Refs.~\cite{golec,jung,zotov,pumplin} suggest the exponential
form of $R_{KT}$:

\begin{equation}\label{rzero}
R' \sim N \exp [- \lambda k^2_{\perp}],
\end{equation}
with $N$ and $\lambda$ in Eq.~(\ref{rzero})
being parameters. However, Eq.~(\ref{rzero}) contradicts to the general requirement of IR stability
(cf. Eq.~(\ref{rsmallk}))  stating that
\begin{equation}\label{rktsmallk}
R_{KT} \sim \left(k^2_{\perp}\right)^{\eta}
\end{equation}
at $k^2_{\perp} \to 0$,
so $R'$ in Eq.~(\ref{rzero}) must be modified. The simplest modification compatible with Eq.~(\ref{rktsmallk}) is

\begin{equation}\label{r}
R_{KT} = N \left(k^2_{\perp}\right)^{\eta} \exp [- \lambda k^2_{\perp} ],
\end{equation}
which agrees with the expressions used in Refs.~\cite{jung,zotov}. So, Eqs,~(\ref{gr},\ref{gb},\ref{r}) present
the complete set of formulas to
describe the non-perturbative input $G_{KT}$ for the gluon distributions in $K_T$ Factorization.
It is interesting to note a further similarity of our expressions in Eqs.~(\ref{gr},\ref{gb})
and the model
suggested in Ref.~\cite{pumplin}, where non-perturbative distributions for the five-quark state are studied.
The quark distributions in Ref.~\cite{pumplin}
involve both a Gaussian (though without the power factor $\left(k^2_{\perp}\right)^{\eta}$)
and  propagators $\sim (k^2_{\perp}/\beta - m^2 + \imath \varepsilon)^{-1}$ of the five
quarks. The latter resembles our factor $Z$ in Eq.~(\ref{zres}) save the difference between
the factors $\imath \varepsilon$ and $\imath \Gamma_{1,2}$.
The quarks in Ref.~\cite{pumplin} are assumed to be free and stable,  the gluon content
is dropped.
In contrast to Ref.~\cite{pumplin}, we do not specify the content of the
spectators
but describe the whole set of spectators through
the resonances and background contributions.

\section{Gluon distributions in Collinear Factorization}

Our next aim is to reduce $D_{KT}$ of Eq.~(\ref{dkt}) to the gluon distribution $D_{col}$
in Collinear Factorization and bring the gluon distribution to the following form:

\begin{equation}\label{dcol}
D_{col} = \int \frac{d \beta}{\beta} D^{(pert)}_{col} (x/ \beta, \mu^2_F) \varphi_{col} (\beta, \mu^2_F),
\end{equation}
with $D^{(pert)}_{col}$ being the perturbative contribution and $\varphi$ being the non-perturbative input; $\mu_F$ is
the factorization scale. To this end,
the $k_{\perp}$-dependence in Eq.~(\ref{dkt}) should be integrated out. At the first sight, such a reduction is impossible:
$D^{(pert)}_{KT}$ and $G_{KT}$ in Eq.~(\ref{dkt}) explicitly depend on $k_{\perp}$, so they both must be integrated and because of that
the integration can yield some entangled mixture
of the perturbative and non-perturbative terms instead of their product as represented in Eq.~(\ref{dcol}).
However, the specific form of the inputs  in Eq.~(\ref{gr},\ref{gb}) allows us to reduce $D_{KT}$ to $D_{col}$
though approximately. According to Eq.~(\ref{dktrb}), $D_{KT}$ consists of $D_{KT}^R$ and $D_{KT}^B$.
Let us consider first the resonance component, $D_{KT}^R$ given by Eq.~(\ref{dktr}). Integration over $\zeta$
therein runs in the resonance region, so only $G_R$ must be integrated:

\begin{eqnarray}\label{intdktr}
D_{KT}^R  &=& \int_x^1 \frac{d \beta}{\beta}
\int_0^{\zeta_0}
\frac{d \zeta}{\zeta} D_{KT}^{(perp)} (x/\beta, q^2/\zeta) G_R (\beta, \zeta)
\\ \nonumber
&\approx& \int_x^1 \frac{d \beta}{\beta}
D_{col}^{(perp)} (x/\beta, \mu'^2_1) \varphi_R (\beta, \mu'^2_1) + O \left(\Gamma'_1/\mu'^2_1\right)
\\ \nonumber
&+& \int_x^1 \frac{d \beta}{\beta}
D_{col}^{(pert)} (x/\beta, \mu'^2_2) \varphi_R (\beta, \mu'^2_2) + O \left(\Gamma'_2/\mu'^2_2\right),
\end{eqnarray}
where $\zeta_0 = w (1 - x/\beta) \approx w$, with $w = 2pq$ as defined in Eq.~(\ref{pq}).
We presume that the resonances are narrow:

\begin{equation}\label{gammamu}
\Gamma'_1/\mu'^2_1 \ll 1, ~~\Gamma'_2/\mu'^2_2 \ll 1.
\end{equation}
In Eq.~(\ref{intdktr}) the notations $D_{col}^{(pert)} (x/\beta, \mu'^2_r)$,  with $r = 1,2$, stand for the perturbative components while
non-perturbative inputs $\varphi_R (\beta, \mu'^2_r)$ are
\begin{equation}\label{phir}
\varphi_R (\beta, \mu'^2_r) = R (\mu^2_r \beta)/\mu^2_r = N \left(\mu^2_r\right)^{\eta -1}\beta^{\eta} e^{- \lambda \beta \mu^2_r}
\approx N_r \beta^{\eta} \left(1 - C_r \beta\right).
\end{equation}

In Eq.~(\ref{phir}) we have denoted $N_r = N \left(\mu^2_r\right)^{\eta -1}$ and $C_r = \lambda \mu^2_r$. Besides, we
presumed that $\lambda \mu^2_r < 1$ to truncate the series of the power expansion of the exponential.
In contrast to $D_{KT}^R$, integration over $k_{\perp}$ (over $\zeta$ in fact) in Eq.~(\ref{dktb}) runs outside the resonance
region and therefore it necessitates
integration of both $G_B$ and $D_{KT}^{(perp)}$. Because of that the outcome scarcely can be represented in the factorized form of Eq.~(\ref{dcol}),
which  violates the reduction to Collinear Factorization.
Fortunately, this violation is small. Indeed,

\begin{equation}\label{intdktb}
D_{KT}^B = \int \frac{d \beta}{\beta}
\frac{d \zeta}{\zeta} D_{KT}^{(perp)} (x/\beta, q^2/\zeta)
G_B (\beta, \zeta) \sim O \left(\Gamma'_1/\mu'^2_1, \Gamma'_2/\mu'^2_2\right),
\end{equation}
so when the resonance input
$G_{KT}^R$ is narrow, i.e. when $\Gamma'_{1,2} \ll \mu'^2_{1,2}$, we can neglect  all contributions
$\sim \Gamma'_1/\mu'^2_1, \Gamma'_2/\mu'^2_2$ in Eqs.~(\ref{intdktr},\ref{intdktb}),
thereby arriving at the following expression for the gluon distribution $D_{col}$ in Collinear Factorization:

\begin{equation}\label{dcol1}
D_{col} \approx \int \frac{d \beta}{\beta} \left[D^{(pert)}_{col} (x/ \beta, \mu'^2_1) \varphi_R (\beta, \mu'^2_1) +
D^{(pert)}_{col} (x/ \beta, \mu'^2_2) \varphi_R (\beta, \mu'^2_2) \right].
\end{equation}
The integrand in Eq.~(\ref{dcol1}) looks similar to the conventional one in Eq.~(\ref{dcol}) but does not
coincide with it. It is easy to bring Eq.~(\ref{dcol1}) to the conventional form (\ref{dcol}) with the perturbative
evolution in the $k_{\perp}$ -space. It can be done in two steps: First, let us introduce
a scale $\mu_F$ so that $|q^2| > \mu_F > \mu'_{1,2}$. Then, using DGLAP (or another evolution approach)
we evolve $\varphi_R (\beta, \mu'^2_1)$ and $\varphi_R (\beta, \mu'^2_2)$
in Eq.~(\ref{dcol1})
from their scales up to the scale $\mu_F$, keeping $\beta$ fixed. This procedure automatically sets
$D^{(pert)}_{col} (x/ \beta, \mu'^2_{1,2})$ in Eq.~(\ref{dcol1}) on the scale $\mu_F$.
As a result, we convert Eq.~(\ref{dcol1}) into Eq.~(\ref{dcol}), where $\varphi_{col}$ is related to the non-perturbative input $\varphi_R$
by the operator $E$ of the DGLAP-evolution:

\begin{equation}\label{phicol}
\varphi_{col} (\beta, \mu^2_F) = E (\mu^2_F, \mu^2_1) \otimes \varphi_R (\beta, \mu^2_1) +
E (\mu^2_F, \mu^2_2) \otimes \varphi_R (\beta, \mu^2_2).
\end{equation}

The evolution operator $E$ in the Mellin (momentum) space is expressed in terms of the DGLAP anomalous dimensions.
In contrast to the non-perturbative inputs $\varphi_R (\beta, \mu'^2_r)$,
the input $\varphi_{col}$ comprises both non-perturbative (through operators $E$) and perturbative (through $\varphi_R$) contributions.

\subsection{Comparison of Eq.~(\ref{phir}) to the standard DGLAP fits}

Let us compare the non-perturbative input $\varphi_R$ with the standard DGLAP fit $\widetilde{\varphi}$.
Quite often (see e.g. Ref.~\cite{fits}) $\widetilde{\varphi}$ is
chosen in the following form:

\begin{equation}\label{fit}
\widetilde{\varphi} (\beta,\mu^2_F) = N  \beta^{-a}(1- \beta)^b (1 + c \beta^d),
\end{equation}
where $N, a,b,c,d$ are phenomenological parameters defined from analysis of
experimental data. All these parameters implicitly
depend on the factorization scale $\mu_F$.  Its value can be chosen arbitrary but factorization convolutions,
where $\widetilde{\varphi}$ participates, do not depend on $\mu_F$.

First of all, let us note that the singular factor $\beta^{-a}$ in Eq.~(\ref{fit}) provides the fast growth of the
parton distributions at small $x$ and leads to their Regge asymptotics.
We proved (see e.g. the overview \cite{egtg1sum}) that the singular factors in the DGLAP fits mimic accounting for the total
resummation of $\ln^n (1/x)$. When such resummation is taken into account, the factors $\beta^{-a}$
become irrelevant and should be dropped. On the other hand, our analysis does not exclude the
factor $\beta^{\eta}$ in Eq.~(\ref{phir}) because $\eta$ is positive.

The factor $(1- \beta)^b$ of Eq.~(\ref{fit}) looks as the large-$x$  asymptotics of
of the parton distributions at $x \to 1$. By this reason, we suggest that the origin of
this factor also is perturbative, so the factor $(1- \beta)^b$ can be excluded from the fit
when the total resummation of $\ln^n (1-x)$ is taken into account.

After the factors $\beta^{-a}$ and $(1- \beta)^b$ have been excluded from Eq.~(\ref{fit}), it
becomes quite similar to the input $\varphi_R$.

\section{Summary}

In the present paper we have performed the detailed study of the general structure of the gluon non-perturbative inputs for
the gluon distributions in hadrons and applied the Resonance Model to construct the gluon inputs.
These inputs can also apply to description of various hadronic high-energy processes, including the DIS structure functions.
We constructed the gluon inputs in Basic Factorization and then calculated the gluon inputs
in the reduced forms so as to use them
in $K_T$- and Collinear Factorizations.

We began with constructing the non-perturbative inputs for the elastic gluon-hadron scattering
amplitudes in the forward kinematics, where we considered the cases of polarized and non-polarized hadrons.
Then, the Optical theorem allowed us to proceed to the  gluon distributions in the hadrons.

Before specifying our model, we derived
general theoretical restrictions (\ref{mir},\ref{muv}) on the non-perturbative inputs,
which are obligatory for any model.
These restrictions follow from the obvious requirement: integrating the factorization convolutions
must yield finite results, i.e. the integrands of the convolutions must be free of IR and UV divergences.

We used those restrictions as criteria for modeling the inputs in
Basic Factorization and reduced them step-by-step to $K_T$- and Collinear Factorizations.
Our model presumes that the gluon non-perturbative inputs in Basic Factorization consist of the polarization structures
and invariant amplitudes
$M_{U,S}$.
For the sake of simplicity, we suggested
in Eq.~(\ref{mrz})
separation
of the $k^2$- and $s_1$- dependence in the expressions for $M_{U,S}$, representing these amplitudes
as
products of the factors $R(k^2)$ and $Z(s_1)$.
Then for specifying the factors $Z (s_1)$
we proposed the Resonance Model
and represented in Eqs.~(\ref{msud},\ref{d}) the inputs through superpositions of the resonances.

Transition from Basic Factorization to $K_T$- Factorization caused reducing the inputs of Eqs.~(\ref{msud},\ref{d})
down to the expressions in Eqs.~(\ref{mkt},\ref{grb}).
Eqs.~(\ref{mkt},\ref{grb}) demonstrate that the inputs in $K_T$-Factorization are again given
by the terms of the resonance type. However, some of the resonances
have maximums  far beyond the region of integration over $k_{\perp}$,
so they can be regarded as a background. Using the models suggested in Refs.~~\cite{golec,jung,zotov,pumplin},
we assumed the exponential (Gaussian) form for the factor $R$ introduced in Eq.~(\ref{mrz}).
Confronting it to Eq.~(\ref{rktsmallk}) following from the requirement of IR stability allowed us to exclude a part
of these models.

Moving from $K_T$- Factorization to Collinear Factorization,
we reduced the expression for the gluon non-perturbative input in Eq.~(\ref{grb}) down to Eq.~(\ref{phir})
and compared it with the conventional expression (\ref{fit}) for
integrated parton distributions.

Finally, we conclude that our calculations prove that the only difference between the
gluon inputs and quark inputs obtained in Ref.~\cite{egtquark} is the difference between their polarization
structures whereas the invariant amplitudes $M_{U,S}$ can be the same. Our study points out that there is a
certain universality between the non-perturbative quark and gluon inputs in each of the available forms of QCD factorization.

\section{Acknowledgements}

We are grateful to A.V.~Efremov, A.~van Hameren, G.I.~Lykasov, W.~Schafer and O.V.~Teryaev for useful discussions.


\begin{thebibliography}{99}


\bibitem{dglap} G.~Altarelli and G.~Parisi, Nucl.~Phys.B126 (1977) 297;
V.N.~Gribov and L.N.~Lipatov, Sov.~J.~Nucl.~Phys. 15 (1972) 438;
L.N.Lipatov, Sov.~J.~Nucl.~Phys. 20 (1972) 95; Yu.L.~Dokshitzer,
Sov.~Phys.~JETP 46 (1977) 641.

\bibitem{egtg1sum} B.I.~Ermolaev, M.~Greco, S.I.~Troyan. Riv.Nuovo Cim. 33 (2010) 57.

\bibitem{colfact}
A.V.~Efremov, A.V.~Radyushkin.  Phys.Lett.B63 (1976) 449,
Teor.Mat.Fiz. 42 (1980) 147,
Theor.Math.Phys.44 (1980)573, Teor.Mat.Fiz.44 (1980)17;
Lett.Nuovo Cim.19 (1977)83;
G.Sterman, S.~Weinberg. Phys. Rev. Lett. 39 (1977) 1436;
S.~Libby, G.~Sterman. Phys. Rev. D18 (1978) 3252; G.~Sterman. Phys.Rev. D17 (1978) 2773, 2789;
D.~Amati, R.~Petronzio, G.~Veneziano. Nucl. Phys. B140 (1978) 54,
Nucl. Phys. B146 (1978) 29;
S.J.~Brodsky and G.P.~Lepage. Phys. Lett. B 87 (1979) 359; Phys. Rev. D 22 (1980) 2157;
J.C. Collins and D.E. Soper. Nucl. Phys.B 193 (1981) 381,
Nucl. Phys.B 194 (1982) 445;
J.C. Collins, D.E. Soper and G.~Sterman. Nucl. Phys.B 250 (1985) 199.
A.V.~Efremov and A.V.~Radyushkin. Report JINR E2-80-521; Mod.Phys.Lett. A24 (2009) 2803;
H.D.~Politzer. Phys. Lett. 70B (1977) 430;
R.K.~Ellis, H.~Georgy, M.~Machacek, H.D.~Politzer, G.G.~Ross. Phys. Lett. 78B (1978) 281;
R.K.~Ellis, H.~Georgy, M.~Machacek, H.D.~Politzer, G.G.~Ross.
Nucl. Phys. B152 (1979) 285.

\bibitem{bfkl} E.A. Kuraev, L.N. Lipatov and V.S. Fadin, Sov. Phys. JETP 44, 443 (1976);
E.A. Kuraev, L.N. Lipatov and V.S. Fadin, Sov. Phys. JETP 45, 199 (1977);
I.I. Balitsky and L.N. Lipatov, Sov. J. Nucl. Phys. 28, 822 (1978).

\bibitem{ktfact} S.~Catani, M.~Ciafaloni, F.~Hautmann. Phys. Lett. B 242 (1990) 97;
Nucl.Phys.B366 (1991) 135.

\bibitem{hefact} J.C.~Collins, R.K.~Ellis. Nucl.Phys. B360 (1991) 3.

\bibitem{fits} G.~Altarelli, R.D.~Ball, S.~Forte and G.~Ridolfi,
Nucl.~Phys.~B496 (1997) 337; Acta Phys. Polon. B29(1998)1145.
E.~Leader, A.V.~Sidorov and D.B.~Stamenov. Phys. Rev. D73 (2006)
034023; J.~Blumlein, H.~Botcher. Nucl. Phys. B636 (2002) 225;
M.~Hirai at al. Phys. Rev. D69 (2004) 054021.

\bibitem{inputmodels} Dmitri Diakonov, V. Petrov, P. Pobylitsa, Maxim V. Polyakov. Nucl.Phys. B480 (1996) 341;
H. Avakian, A.V. Efremov, P. Schweitzer, F. Yuan. Phys.Rev. D81 (2010) 074035;
Ivan Vitev, Leonard Gamberg, Zhongbo Kang, Hongxi Xing.  PoS QCDEV2015 (2015) 045;
Asmita Mukherjee, Sreeraj Nair, Vikash Kumar Ojha. Phys.Rev. D91 (2015), 054018.

\bibitem{pasquini} C. Lorce, B. Pasquini, P. Schweitzer. JHEP 1501 (2015) 10.


\bibitem{golec} K. Golec-Biernat, M. Wusthoff, Phys. Rev. D60 (1999) 114023.

\bibitem{jung} H.~Jung hep-ph/0411287.

\bibitem{zotov} A.V.~Lipatov, G.I.~Lykasov, N.P,~Zotov. Phys. Rev. D59.
(2014) 014001;
A. A.~Grinyuk, A. V.~Lipatov,~G. I. Lykasov and N. P.~Zotov. Phys. Rev. D 93 (2016) 014035.

\bibitem{pumplin} Jon Pumplin PRD 73 (2006) 114015.

\bibitem{brod} S. J.~Brodsky, P. Hoyer, C. Peterson, N. Sakai, Phys. Lett. B93 (1980)451.


\bibitem{lattice} Yan-Qing Ma, Jian-Wei Qiu. Int.J.Mod.Phys.Conf.Ser. 37 (2015) 1560041;
Yan-Qing Ma, Jian-Wei Qiu. 	arXiv:1404.6860;
Martha Constantinou. PoS CD15 (2015) 009.

\bibitem{sud} V.V.~Sudakov. Sov. Phys. JETP 3(1956)65.

\bibitem{egtfact} B.I.~Ermolaev, M.~Greco, S.I.~Troyan. Eur.Phys.J. C71 (2011) 1750;
B.I.~Ermolaev, M.~Greco, S.I.~Troyan.  Eur.Phys.J. C72 (2012) 1953.

\bibitem{egtquark} B.I.~Ermolaev, M.~Greco, S.I.~Troyan. Eur.Phys.J. C75 (2015)7, 306.














\end{thebibliography}
\end{document}